# SCALE EFFECTS METHODOLOGY APPLIED TO UPRISING JETS IN ASTRID

**B.Jourdy[1,3], D.Guenadou[1], N.Seiler[2], A.Labergue[3], M.Gradeck[3]**
[1] CEA Cadarache, DES/IRESNE/DTN/STCP/LTHC
[2] CEA Cadarache, DES/IRESNE/DTN/SMTA/LMAG
13115 Saint-Paul-Lez-Durance, FRANCE
[3] Université de Lorraine, CNRS, LEMTA, F-54000 Nancy, France
benjamin.jourdy@cea.fr; david.guenadou@cea.fr; nathalie.seiler@cea.fr;
alexandre.labergue@univ-lorraine.fr; michel.gradeck@univ-lorraine.fr

**ABSTRACT**

CEA is involved in the development of the 4th generation of nuclear reactors, among which are the Sodium Cooled Fast Reactors (SFR). To support their design and safety, specific codes were developed and should be validated using experimental results from relevant mock-ups. Due to the complexity of building a full-sized prototype in the nuclear field, most of the experiments are performed on reduced-sized models but it may lead to scale effects. Such scale effects are under study in this paper, focusing on a critical issue in SFR reactor, which is the rising of the jet outgoing the core at low power.

For this purpose, a scale effects methodology is detailed, using the SFR's reduced scale mock-up MICAS as the reference scale. To achieve that, dimensionless Navier-Stokes equations under Boussinesq's approximation are considered and the Vaschy-Buckingham theorem is applied to determine the most relevant dimensionless numbers, among them the densimetric Froude number. Experimental campaigns have been performed to measure velocity fields using the Particle Image Velocimetry (PIV). Their analysis clearly shown dependency of mean pathway of the jet on this dimensionless number. Particularly, plotting the jet angle as a function of the normalized densimetric Froude number, a change of behaviour for a threshold value of 0.45 has been observed. This result also shown negligible effects on the jet rise happening before jet impingement on the Upper Core Structure (UCS).

## 1. INTRODUCTION

In order to close the nuclear fuel cycle and use the wide spread uranium 238 isotope instead of the uranium 235 currently being used in Gen II and Gen III nuclear reactors, CEA is involved in the development of Gen IV sodium-cooled fast neutron reactors (SFR). Building an experiment at scale 1 for characterizing the reactor prototype behaviour is complex and expensive, that is why most experiments are performed on small-scale mock-ups to assess new design options and validate associated simulation softwares. Due to sodium opacity and its reactivity with water, sodium experiments are very complex to carry out, even at small scale. That is why most of such experiments are performed with a simulating fluid which is water, owing to its close properties to the sodium ones.

In 2015, in order to support the French experimental reactor project called ASTRID [1], the 1:6th scale mock-up MICAS was commissioned to study the thermal-hydraulics behaviours inside the upper plenum of ASTRID. This mock-up allows investigations of different thermal-hydraulics issues as the gas entrainment at free surface [2] or the behaviour of the jets coming out from the core [3]. For this specific issue, these jets can be subjected to thermal-hydraulics oscillations at low power operation. And, after impinging the Upper Core Structure (UCS), they may raise due to buoyancy forces and lead to oscillating temperature of some submerged vessel components and thus induce thermal stress, which could have some consequences on the lifetime of the reactor.

This study focuses on the thermal-hydraulic behaviour of the hot impinging jets out of the core, especially on the transition from inertia-dominated to buoyancy-dominated jets. The first part briefly introduces ASTRID prototype and MICAS mock-up, and provides details about the phenomenon under investigation. A theoretical study is also provided. Then, a description of the experiments conducted on MICAS and their experimental results are discussed. These results will be used as a reference for a later scale effect analysis by comparison with results at lower scales.

## 2. THE ASTRID PROTOTYPE AND MICAS MOCK-UP

### 2.1. General Overview

From 2011 to 2019, the CEA has been involved in the development of the 4th generation ASTRID (Advanced Sodium Technological Reactor for Industrial Demonstration) prototype [4]. This project aimed at demonstrating the technical options chosen and at estimating the operating costs. Its power was 600MWe. However, even if this project has stopped in 2019, the CEA still carries on studies on sodium reactors and associated technologies. Especially, efforts head to build a numerical reactor, implying a strong validation of numerical codes.

The MICAS facility is a 1:6th sized mock-up representing the hot plenum of the ASTRID reactor, based in the CEA Cadarache [5]. The scale was chosen to be a compromise between the overall size and the detail of the geometry of the vessel as MICAS geometry is homothetic to ASTRID's one, but some geometrical simplifications were necessary. Its dimensions are about 2.5 m in diameter and 1.7 m in height. Cut views of both ASTRID and MICAS are shown in Figure 1.

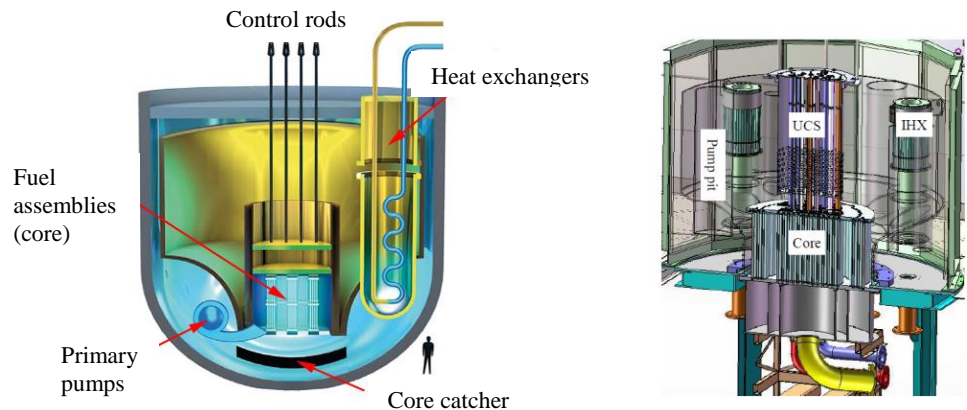

**Figure 1. Cut Views of ASTRID (Left) and MICAS (Right)**

### 2.2. Phenomenon of Interest and Problematic

Figure 2 represents a simplified scheme of the flow above the core. The left part represents the flow at low mass flow rate, the right represents the flow at nominal flow rate conditions. Upstream of the core, 288 hot jets are first impinging and going through a porous plate, inducing a pressure loss. This porous plate deviates a small part of the flow horizontally. Part of the water (around 15% [3]) flows through the Upper Core Structure (UCS) and exits around the sheath tubes. The bottom plate of the UCS deflects the entire part of the flow, resulting in an almost unique radial jet spreading into the upper plenum of vessel where the fluid temperature is lower than the jet's one.

At low power operating conditions, the inertia of this jet decreases and buoyancy effects become predominant and the jet can rise under some conditions. If so, the flow pattern in the vessel is modified and leads to thermal fluctuations, causing thermal stress of the components of the vessel. As this phenomenon is crucial for the life expectancy of the reactor, a deep investigation is necessary to clearly identify the conditions for this transition.

Experimental results on MICAS are used to validate numerical codes to simulate ASTRID's thermal-hydraulics behavior. However, as we are using a downscaled mock-up, we need to ensure that the experimental conditions are representative of the reactor ones and that the phenomenon under investigation (i.e the raise of the impinging jets) can be transposed from a scale to another without too much distortions. So, the main issue of this study is to find the most relevant dimensionless numbers to ensure the similarity of the flow; moreover, we should also study the dependence of the raise of the jet on relevant dimensionless numbers.

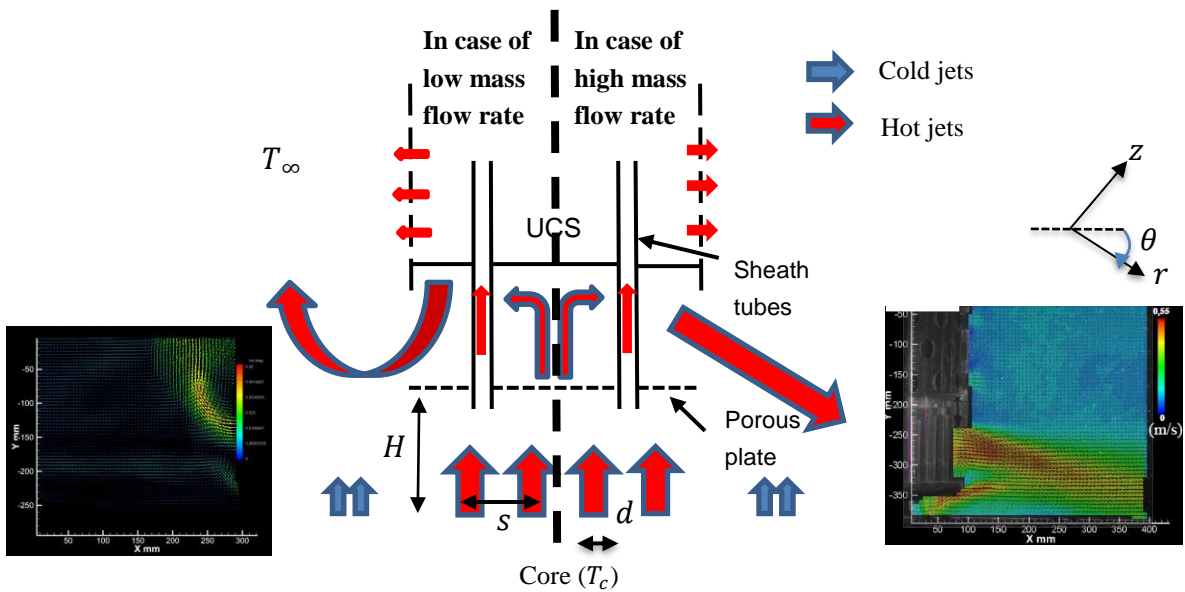

**Figure 2. Scheme of the flow path above the core in ASTRID and MICAS in
two different cases of mass flow rate and associated experimental results
(velocity vectors colored according to velocity scale)**

## 3.  THEORETICAL STUDY

### 3.1.  Similarity of the Flow

In order to be representative of the ASTRID's flow behavior in MICAS, we applied a dimensional analysis [6] to MICAS that led to the dimensionless numbers shown in the Table 1. This allows to select the ones whose conservation ensures both the similarity of the flow and heat transfer between two different scales (a nomenclature is given at the end of the document). This analysis also leads to geometrical dimensionless parameters such as $\pi_1$ or $\pi_9$ : these parameters do not interfere from ASTRID to MICAS as the transformation is linear from a scale to another, but can be important for the design of mock-up that need geometrical distortion.

**Table 1. Dimensionless numbers obtain with Vaschy-Buckingham theorem**

| $\pi_1 = \dfrac{H}{L}$ | $\pi_2 = \dfrac{p}{\rho.u^2} = Eu$ | $\pi_3 = \dfrac{\eta}{\rho.u.L} = \dfrac{1}{Re}$ |
|---|---|---|
| $\pi_4 = \dfrac{g.L}{u^2} = \dfrac{1}{Fr^2}$ | $\pi_5 = \dfrac{T - T_\infty}{T_c - T_\infty}$ | $\pi_6 = \beta(T_c - T_\infty)$ |
| $\pi_7 = \dfrac{\alpha}{u.L} = \dfrac{1}{Pe}$ | $\pi_8 = \dfrac{\rho_\infty - \rho}{\rho_\infty}$ | $\pi_9 = \dfrac{s}{L}$ |

However, as the conservation of every $\pi$ groups is not possible from a scale to another, we use the dimensionless governing equations known as the conservation equations in order to determine the most relevant dimensionless numbers and calculate our experimental conditions. The study of the MICAS mock-up can be divided in two different zones: first, the under-UCS zone where 288 hot jets exit the core, and then the vessel zone, where a unique jet, coming from the merge of the 288 core jets after their impact on the UCS bottom plate, is diffusing into the vessel.

Just above the core, we assume each jet to be an axisymmetric incompressible steady jet where the velocity vector $u$ in the $(r,z)$ cylindrical coordinate system is defined by $U$ its axial component and V its radial component as $\vec{u} = U \vec{e_z} + V \vec{e_r}$. In this zone, the axial component is vertical as shown Figure 2. The scaling parameters are chosen as $U_0$, the average axial velocity at the jet exit, $d$ the jet diameter, $T_c$ the jet's temperature out of the core and $\rho_\infty$ the environment temperature, equal to the jet's one in this zone. The notation $X^*$ means the dimensionless parameter $X/X_0$, $X_0$ being the scaling parameter associated to the parameter $X$. This leads to the following dimensionless conservation equations:

Mass conservation:

$$\frac{\partial U^*}{\partial z^*} + \frac{1}{r^*}\frac{\partial}{\partial r^*}(r^* V^*) = 0 \tag{1}$$

Energy conservation:

$$\left(U^* \frac{\partial T^*}{\partial z^*} + \frac{V^*}{r^*}\frac{\partial T^*}{\partial r^*}\right) = \frac{1}{Re}\frac{1}{Pr}\frac{1}{r^*}\frac{\partial}{\partial r^*}\left(r^* \frac{\partial T^*}{\partial r^*}\right) \tag{2}$$

Between the porous plate and the UCS, the overall fluid temperature is considered homogeneous being the hot jets' temperature, around 55°C. The dimensionless momentum equation for a jet in this zone can be reduced to its axial one, as $V \ll U$, and is given by equation (3):

$$U^* \frac{\partial U^*}{\partial z^*} + V^* \frac{\partial U^*}{\partial r^*} = -\frac{1}{\rho^*} \frac{\partial Eu}{\partial z^*} + \frac{1}{Re} \frac{\partial}{r^* \partial r^*} \left\{ r^* \frac{\partial U^*}{\partial r^*} \right\} \quad (3)$$

In this zone, the similarity is ensured by the conservation of the dimensionless numbers from equation (2) and (3), and the conservation of the interaction between jets. Important parameter to ensure that the dynamic interaction between the 288 jets is preserved, and thus that we can consider only one single jet, are the dimensionless numbers $\pi_1$ and $\pi_9$ (see Table 1) with $H$ the nozzle-to-plate distance, $d$ the jet diameter and $s$ the inter-jet distance [7]. As MICAS is a linear downscaled mock-up from ASTRID, these ratios are automatically preserved. Thus, the main dimensionless numbers of interest in this zone are the Euler number and the Reynolds number. As this zone has an almost constant temperature, the Prandtl number is not important under the UCS.

After impingement of the UCS, the jets are merging and a single hot jet is spreading in the vessel where the surrounding fluid is colder. As $(\rho_\infty - \rho)/\rho_\infty \sim 0.2\%$, we assume the Boussinesq approximation to be valid in this zone. As pressure variations are low in a free submerged jet, we assume the pressure terms to be negligible and we keep the same scaling parameters. The flow is still axisymmetric but as the flow is deviated by the impact of the UCS bottom plate, the radial component of the momentum equation cannot be neglected. Thus, an angle $\theta$ from the non-perpendicular deviation of the jet is expressed as a projection of the gravity term onto the $(\vec{r}, \vec{z})$ coordinates (Figure 2), leading to the dimensionless momentum equations in this zone given equation (4):

$$\begin{cases} U^* \frac{\partial U^*}{\partial z^*} + V^* \frac{\partial U^*}{\partial r^*} = \frac{1}{Re} \left\{ \frac{\partial}{r^* \partial r^*} \left( r^* \frac{\partial U^*}{\partial r^*} \right) + \frac{\partial^2 U^*}{\partial z^{*2}} \right\} - \left( \frac{1}{Fr^2} + \frac{1}{Fr_D^2} \right) \cos(\theta) \\ U^* \frac{\partial V^*}{\partial z^*} + V^* \frac{\partial V^*}{\partial r^*} = \frac{1}{Re} \left\{ \frac{\partial}{r^* \partial r^*} \left( r^* \frac{\partial V^*}{\partial r^*} \right) + \frac{\partial^2 V^*}{\partial z^{*2}} \right\} - \left( \frac{1}{Fr^2} + \frac{1}{Fr_D^2} \right) \sin(\theta) \end{cases} \quad (4)$$

As the flow is fully turbulent in MICAS ($Re_d > 10^4$ for each jet at the core exit), the Reynolds number may have a small influence on the jet's behavior as jets tends to be Reynolds invariant [8]. The main parameters to achieve similarity from ASTRID to MICAS (as long as the flow is fully turbulent) are the Euler's number for the jets under the UCS, and the densimetric Froude number for the jet spreading in the vessel.

The later dimensionless number is defined by equation (5) where $\Delta \rho = \rho_\infty - \rho$ is the density difference between the hot jets and the surrounding environment, $u$ is the velocity of the jet, $g$ the gravitational constant and $L$ the characteristic length scale of the jet, i.e. the jet diameter at the core exit.

$$Fr_D = \frac{u}{\sqrt{(\Delta \rho / \rho_\infty) g L}} \quad (5)$$

This densimetric Froude number is frequently used for buoyant jet studies under Boussinesq approxiation [7] [9] [10], giving credit to its choice as a similarity parameter. For a jet, high densimetric Froude numbers (such as $Fr_D \gg 1$, i.e. in order of magnitude $Fr_D > 10$) describe momentum-driven flows. When its value decrease, buoyancy effects become more important until the flow becomes buoyancy-driven when $Fr_D \to 0$.

## 3.2. Buoyancy in Jets

For fully turbulent submerged free jets under Boussinesq's approximation, the buoyancy problem only depends on the initial jet's angle $\theta_0$, and the initial jets parameters such as the flow rate $Q_0$, buoyancy $B_0$ and momentum $M_0$ [10]. $S_0$ is the initial exit surface of the jet. These variables are defined as (equation (6)):

$$\begin{cases} Q_0 = u_0 S_0 \\ M_0 = u_0 Q_0 \\ B_0 = Q_0 g \dfrac{\rho_\infty - \rho_0}{\rho_\infty} \end{cases} \quad (6)$$

Here, two characteristics length scales can be defined as $l_M = M_0^{3/4} / B_0^{1/2}$ and $l_Q = Q_0 / M_0^{1/2}$. $l_M$ represents the characteristic length from momentum-driven to buoyancy-driven jet, and $l_Q$ represents the characteristic length at which the jet's exit still influences the flow. For a single round jet, these length scales are related to the jet's characteristic length scale $L$ by the relation $S_0 = \pi L^2 / 4$, the characteristic length scale of a round jet being its diameter.

They are also related to the densimetric Froude number (equation (7)) as:

$$\frac{l_Q}{l_M} = \frac{K_0}{Fr_D} \quad (7)$$

$K_0$ is a constant that depends on the jet's nozzle. For a round jet, $K_0 = (\pi/4)^{1/4}$. For high $Fr_D$, it has been demonstrated [10] that the geometrical parameters such as the distance from the nozzle exit to the onset of its rise $X_Z$ as shown in Figure 3 only depends on the characteristic length scale, the densimetric Froude number and the initial angle:

$$\frac{X_Z}{L\, Fr_D} = K(\theta_0) \quad (8)$$

In our case, the characteristic length scale is the jet diameter. $K(\theta_0)$ is a constant that only depends on the initial jet's angle. This relation shows the dependency of the jet's behaviour on the densimetric Froude number. In this study, we characterize the rise of the jet by two parameters: the distance before its rise $Xz$ and the final angle $\theta_f$ as presented in Figure 3 below. This study will only focus on the evolution of this final angle with the densimetric Froude number.

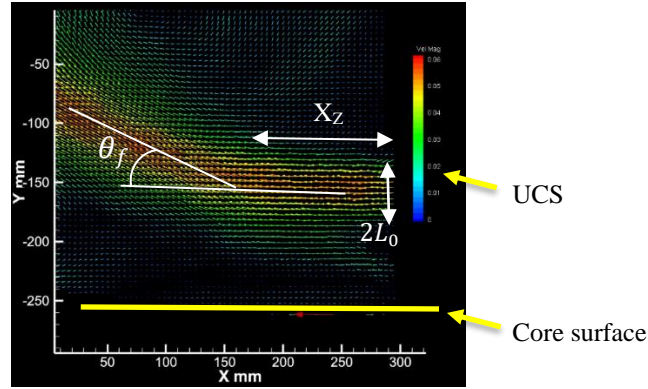

**Figure 3. Characteristics features of the jet's rise**

## 4. EXPERIMENTAL SETUP

### 4.1. Experimental Conditions

The inlet conditions of MICAS taken to be representative of the ASTRID reactor are calculated based on the similitude in densimetric Froude number at the core exit. The similitude in this zone allows the control of all parameters (velocity and temperature) and a relation between the deviated jet's behaviour and the inlet conditions may then be determined. The influence of the Euler number on the behaviour of the jet will be studied later. In order to observe the raise of the jet, we decrease the inlet velocity by decreasing the flow rate and we keep the temperature differences between the jets and the environment, leading to a decrease of the densimetric Froude number due to the decrease of the mass flow. The hot and cold mass flow rates are measured with Coriolis flowmeters ($\pm 0.1\%$ of measurement range).

Most of the components in MICAS are made of optical grade transparent polymer to allow laser measurements. The maximal temperature condition is 60°C and the minimal temperature which can be reached is around 10°C for the cold jets in the vessel. As we keep the same repartition of the flow in MICAS as in ASTRID (95% of hot jets, 5% of cold ones), the operating conditions are given in Table 2. Hot jets have a temperature of 55°C and cold ones are at 10°C [5].

**Table 2. Experimental conditions on MICAS compared to ASTRID's ones**

|  | Experimental conditions | |
|---|---|---|
|  | ASTRID (nominal) | MICAS (20% to 100% of nominal condition) |
| $T_c$ (°C) | ~570 | ~55 |
| $T_c - T_\infty$ (°C) | ~25 | 1,2 to 4,4 |
| $Q_c$ ($m^3 \cdot h^{-1}$) | ~32 214 | 32 to 165 |
| $Fr_D$ | ~ 3 to 27 | |

## 4.2. The Velocity Measurement Setup

The velocity fields are measured thanks to Particle Image Velocimetry (PIV) as shown in Figure 4, allowing us to record 2D large velocity fields (300x300mm) of the jets after impingement. In our setup, the laser source is set on a 3-axis motion table to get a positioning as accurate as possible. Nylon particle with a density of 1000 kg.m$^{-3}$ are mixed with the water. These particles with a 4µm diameter are enlightened by the laser sheet and scatter the light. Then, their motion is recorded by a 4 Mpixels CCD camera, perpendicular to the laser sheet. Its acquisition rate is 15 Hz and velocity fields are averaged over 150 images, i.e. 10s of acquisition, and post-processed with the software INSIGHT 4G (version 11.1.0.5) from TSI. PIV results averaged on a higher number of images (over 300) gave identical results, validating the stationary nature of the flow for a 10s acquisition.

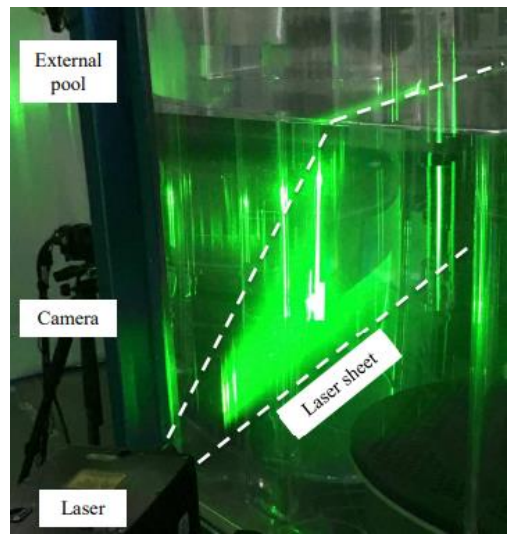

**Figure 4. PIV setup on MICAS**

## 4.3. Temperature Measurements

The inlet temperature for both hot and cold jets are measured with PT100 probes ($\pm 0.1°C$). We also put PT100 probes in the UCS to ensure that we have no thermal exchanges between the hot jets and its environment before their impingement on the UCS and spreading into the vessel.
In the vessel, we use PT100 probes to measure the ambient temperature. These probes are disposed in the middle of the vessel and next to the pump pit (green dots in the Figure 5), to measure a mixing temperature. We also use an array of thermocouples ($\pm 1°C$) to have informations about the stratification and the establishment of the flow's temperature. These 15 thermocouples are disposed on a pole with spacing of 1cm.

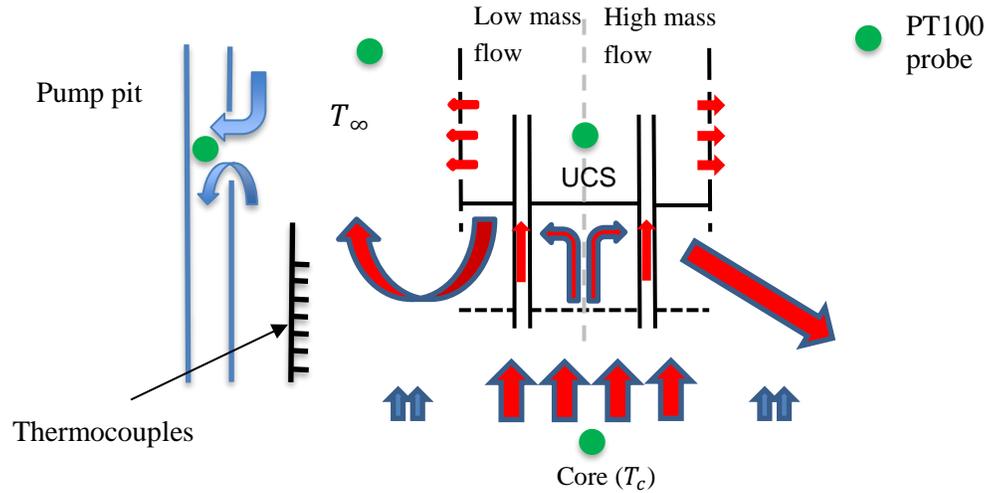

**Figure 5. Temperature measurements in the vessel in MICAS**

## 5. EXPERIMENTAL RESULTS AND COMMENTS

### 5.1. Velocity Measurement Results

The time-averaged PIV results show the raise of the jet in the vessel (displayed in Figure 6) for four typical injection conditions. For high mass flow rate (i.e. momentum-driven flow), we notice an initial angle around 20° (picture 1 of the Figure 6). This angle seems to be constant when buoyancy becomes less important and only depends on the geometry of the UCS. When the mass flow rate decreases, we notice in picture 2 of the Figure 6 that the jet starts to rise and so in pictures 3 and 4, with an increasing final angle.

Even if the parameter that we modified from an experiment to another one is the inlet flow rate being decreased (decreasing the inlet velocity and so the densimetric Froude number as shown in equation (5)), this modification can induce variations of temperature between two experiments where the temperature difference was supposed to be constant ($T_c - T_\infty \sim 2°C$). This leads to cases as shown in pictures 3 and 4 in the Figure 6 where, for a higher flow rate (i.e. a higher Froude number), the angle in picture 4 is still higher because the temperature difference is higher too, leading to a lower densimetric Froude number. This example highlights the difference between the Froude number $Fr$ and the densimetric Froude number $Fr_D$ in the angle study.

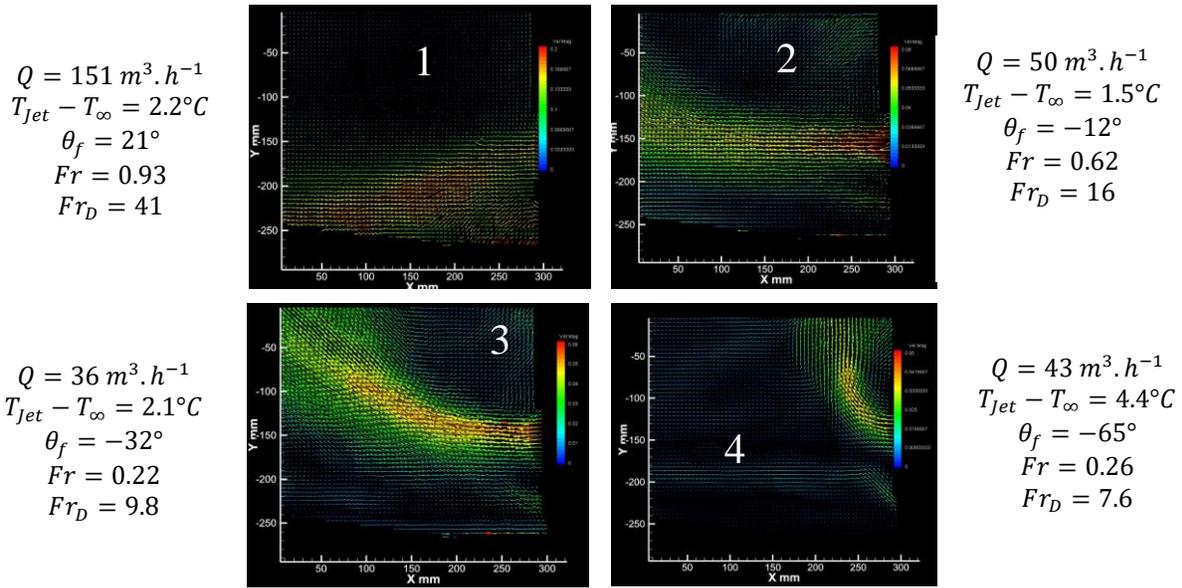

$Q = 151\ m^3.h^{-1}$
$T_{Jet} - T_\infty = 2.2°C$
$\theta_f = 21°$
$Fr = 0.93$
$Fr_D = 41$

$Q = 50\ m^3.h^{-1}$
$T_{Jet} - T_\infty = 1.5°C$
$\theta_f = -12°$
$Fr = 0.62$
$Fr_D = 16$

$Q = 36\ m^3.h^{-1}$
$T_{Jet} - T_\infty = 2.1°C$
$\theta_f = -32°$
$Fr = 0.22$
$Fr_D = 9.8$

$Q = 43\ m^3.h^{-1}$
$T_{Jet} - T_\infty = 4.4°C$
$\theta_f = -65°$
$Fr = 0.26$
$Fr_D = 7.6$

**Figure 6. PIV results with decreasing mass flow**

### 5.2. Temperature Results

As expected, the ambient temperature in the upper plenum of the vessel is subjected to thermal fluctuations. We started measurements when the flow was stabilized, i.e. when the temperature in the vessel was fluctuating around its mean value as shown in Figure 7. When the flow was steady, we processed to PIV measurements. The fluctuations around the mean value in the vessel are increasing with the mass flow. Maximum fluctuations observed go up to $0.4°C$ around the mean value for $Q = 151\ m^3.h^{-1}$ (see the Figure 7), with an uncertainty from PT100 probes of $\pm 0.1°C$.

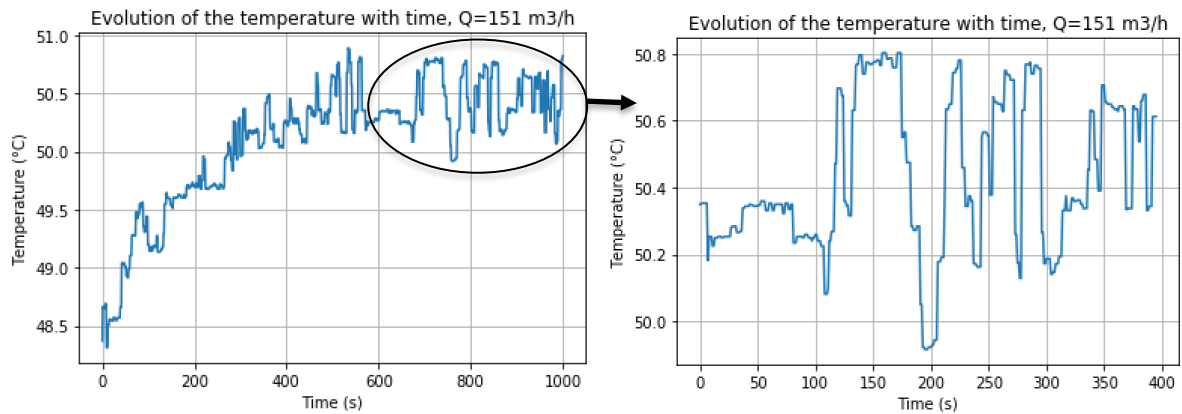

**Figure 7. Stabilization of the temperature in the vessel with time for Q = 151 m³/h measured with PT100 probe**

## 5.3. Angle of the Jet

### 5.3.1. Determination of the jet's angle and width

We define the final angle of the jet being the angle between the horizontal (referred as 0°) and jet centreline at the point where the jet start to rise. This angle is calculated with PIV results by plotting the maximum velocity magnitudes in the jet (representing the jet centreline), and defining a window as shown Figure 8 in which we make a linear regression on the maximal magnitudes. This window is user-defined as it has to differ from a test to another, the jet being in a different location in each case as shown Figure 6. A fixed window would lead to aberrant points if too far from the centre of the jet. The window cannot either be the size of the PIV image, as it would catch the initial angle in the linear regression. Hence, the window dimensions only need to catch the jet centreline after the jet rise (i.e. after the distance $X_Z$ from the inlet), where $\theta \rightarrow \theta_f$. We checked that no modifications appear while moving the window or modifying its size in this zone. This angle is thus an approximation, valid when the jet's trajectory has a clear transition between the initial angle $\theta_0$ and the final angle $\theta_f$ as shown Figure 8, but with an increased uncertainty when the jet describes an arc as shown Figure 6 picture 4. Hence, this linear regression is considered valid in our study as we will mainly focus on the transition when $\theta_f = 0°$. The width of the radial jet is calculated by plotting the velocity magnitudes at the outlet: a velocity vector belongs to the jet if its value is superior to 10% of the maximum velocity on this profile.

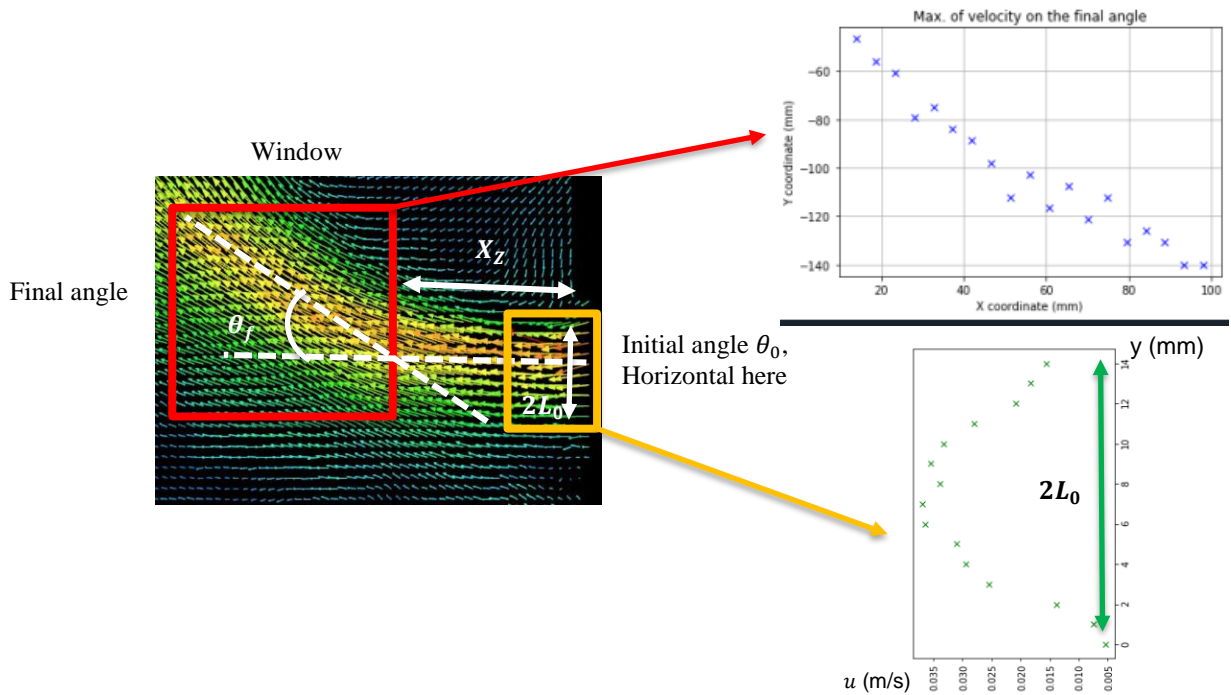

Figure 8. Calculation of the final angle and the width of the radial jet after impingement of the UCS

### 5.3.2. Final angle results

In order to show the jet's behaviour dependency on the densimetric Froude number, we plot the evolution of the final angle as a function of this dimensionless number. The $Fr_D$ is calculated in two different zones:

in the entrance of the vessel after impingement of the UCS, and at the core's exit. In the vessel, equations (4) and (8) shown a dependency of the jet's behaviour only on the $Fr_D$, result we want to confirm in our experiment. But, as we only control inlet conditions out of the core, we need to study the evolution of this angle with the $Fr_D$ calculated with our inlet conditions. In this zone, the jet's behaviour is supposed to be dependent on the Euler number as shown by equation (3), before diffusing in the vessel and be dependent on the $Fr_D$.

First, we want to ensure that the behaviour of the free submerged jet in the vessel after impingement of the UCS bottom plate only depends on the densimetric Froude number. To do so, we calculate the experimental densimetric Froude number regarding the vessel features: the characteristic length is the jet's half-width $L_0$, calculated thanks to PIV results. The velocity is an output velocity, i.e. an average of the velocity vectors on the jet width $2L_0$ (equation (9)). The density differences are given thanks to the PT100 probes' measurements and tabulated values of temperature dependant density.

$$u_{Vessel} = \frac{1}{n_{vec}} \sum_{0}^{2L_0} u_i \qquad (9)$$

Figure 9 shows that the evolution of $\theta_f$ seems to be linear with the $Fr_D$ on a range from 0.5 to 4.7, beyond which the angle becomes constant (between 17° and 21°). After this threshold value of $Fr_D = 4.7$, buoyancy effects become negligible and the jet behave like a momentum-driven jet only. A later study of the measurement uncertainties will provide more information about the linearity obtained in these preliminary results.

This asymptotic behaviour at high $Fr_D$ (i.e. at high mass flow rate and / or low density difference) may be dependent on the UCS geometry, such hypothesis will be studied later. A linear regression on the linear zone gives a critical densimetric Froude number of about 2.5, defined as the transition from positive to negative angle. This figure shows the dependence of the angle on the densimetric Froude number as expected by equations (4) and (8).

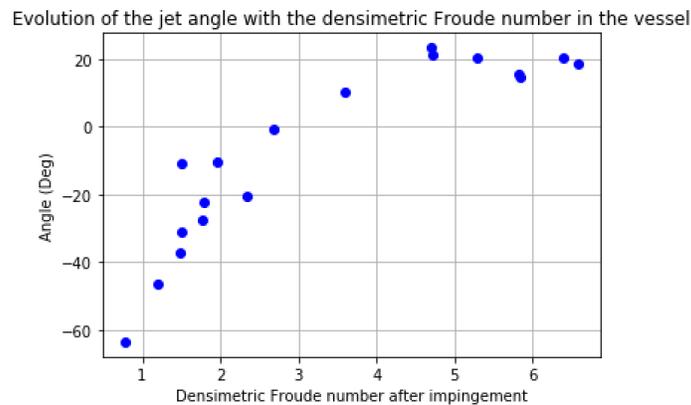

**Figure 9. Evolution of the final angle with the densimetric Froude number in the upper plenum of the vessel**

However, the input parameters in MICAS are the temperature and flow rate at the core exit, where we applied the conservation of the densimetric Froude number between ASTRID and MICAS. As we decrease the inlet flow rate to decrease the inlet densimetric Froude number, we evaluate the behaviour of the jet with these inlet conditions.

Figure 10 represents the evolution of the final angle $\theta_f$ with the initial densimetric Froude number $Fr_D$ using the inlet conditions at the exit of the core as input parameters. This $Fr_D$ is calculated with the same differences of density as previously (the jet's and ambient temperature being respectively the same in both cases), but the characteristic length is here the half-diameter of the jets in the core measured with PIV results as shown Figure 8, and the velocity is an averaged outlet velocity of the jets given equation (10).

$$u_{Core} = \frac{Q}{n_{jet}\, S_{jet}} \tag{10}$$

The evolution of $\theta_f$ seems again to be linear with the $Fr_D$ on a range from 7 to 36 before reaching the asymptotic value around 20°. Buoyancy become negligible for $Fr_D = 36$. A linear regression on the linear zone gives a critical densimetric Froude number of 20.

As the jets undergo non-linear phenomena such as loss of pressure induced by the porous plate, deviation from the sheath tubes or impingement of the UCS, the evolution of the final angle was not expected to be similar to the evolution observed in Figure 9. To enlighten the differences due to non-linear phenomena happening between the core and the UCS, we normalize these results with the densimetric Froude number obtained at nominal mass flow rate as shown equation (11) in both cases (i.e. at 100% of the flow rate as defined Table 2).

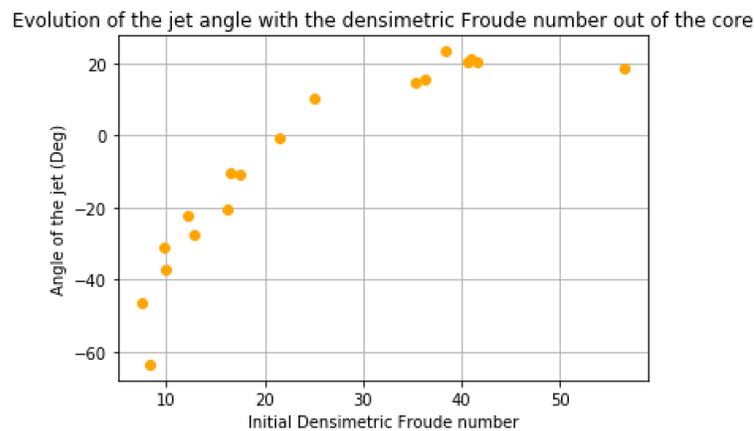

**Figure 10. Evolution of the final angle with the densimetric
Froude number at the outlet of the core**

$$Fr_D^* = \frac{Fr_D}{Fr_{D\,100\%}} \tag{11}$$

Figure 11 gather the experimental data from Figure 9 and Figure 10 normalized by the densimetric Froude number at nominal mass flow rate as shown equation (11). For each plot (Figure 9 and Figure 10), the densimetric Froude number is being normalized by its own nominal value (and not by a unique value common to both experiments), as parameters such as diameter or velocities are not calculated the same way. The aim is to see if, while normalized, the problem of a group of jets flowing through a porous plate and impinging a structure can be compared to a simple jet flowing into a colder environment. This figure shows that the results are close for the two experiments, leading to the conclusion that non-linear phenomena have not many influence on the raise of the jet in the upper plenum at MICAS's scale. However, this affirmation has to be confirmed with lower scales studies.

The normalized critical densimetric Froude number for which the jet's angle is $\theta_f = 0°$ seems to be between $Fr_D = 0.41$ and $Fr_D = 0.45$. The transition from buoyant to momentum-driven jet is reached for $Fr_D \approx 0.75$.

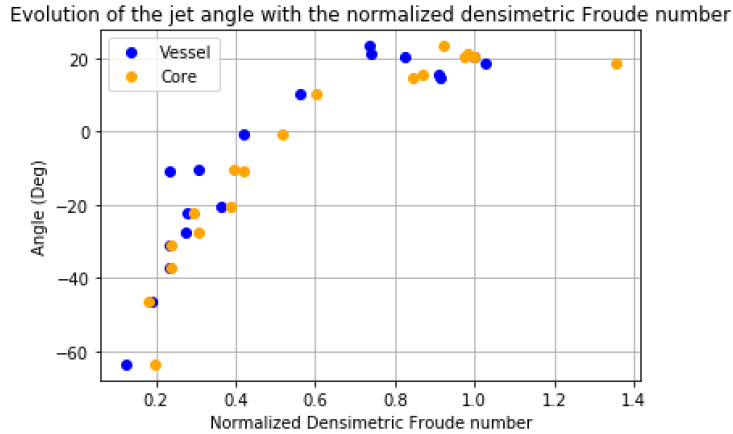

**Figure 11. Evolution of the final angle with the normalized densimetric Froude number**

## 6. CONCLUSIONS

In order to be representative of the flow and heat transfers in ASTRID, experimental inlet conditions on the 1:6 scaled mock-up MICAS are derived from densimetric Froude number conservation, applied to the inlet conditions out of the core. The phenomenon under study being the rise of a hot jet in a colder environment, experiments have been conducted in MICAS. Theoretical study implies that the behavior of the jet depends on the Euler number under the UCS and on the densimetric Froude number in the vessel. Thanks to PIV and temperature measurements, we have been able to plot the evolution of the final angle with the densimetric Froude number, this number being calculated in two different zones, i.e. out of the core and in the vessel after impingement of the UCS. In both cases, the evolution of the final angle seems to be linear with the densimetric Froude number until the jet becomes momentum-dominated only. When buoyancy effects become negligible, the jet's angle keeps a constant value around 20°, that is assumed to be dependent on the mock-up geometry. Differences were expected between these two results, as non-linear phenomena appear between the exit of the core and the spreading in the vessel. But normalized results shown that these non-linear phenomena were negligible on the evolution of the final angle. This first result could imply that phenomena occurring under the UCS such as the loss of pressure induced by the porous plate or the sheath tubes have negligible effects on the rise of the jet inside the upper plenum. The critical normalized densimetric Froude number for which we consider that the jet rises has a value of 0.45. A comparison of this result with the critical densimetric Froude number at other scales will provide information about scale effects on this phenomenon.

## NOMENCLATURE

| | | | |
|---|---|---|---|
| $B_0$ | Buoyancy flux $[m^4.s^{-3}]$ | *Greek symbols:* | |
| $d$ | Jets diameter $[m]$ | $\alpha$ | Thermal diffusivity coefficient $[m^2.s^{-1}]$ |
| $g$ | Gravity acceleration $[m.s^{-2}]$ | $\beta$ | Thermal expansion coefficient $[K^{-1}]$ |
| $g'$ | Reduced gravity $[m.s^{-2}]$ | $\theta$ | Angle of the jet with the horizontal $[deg]$ |
| $H$ | Nozzle-to-plate distance $[m]$ | $\eta$ | Dynamic viscosity $[kg.m^{-1}.s^{-1}]$ |
| $K$ | Constant | $\rho$ | Density of the jets $[kg.m^{-3}]$ |
| $L$ | Characteristic length scale $[m]$ | | |
| $l_M$ | Length scale from momentum to buoyancy driven jet $[m]$ | *Subscripts*: | |
| | | $\infty$ | Ambient environment |
| $l_Q$ | Length scale at which the jet's exit still influence the flow $[m]$ | c | Core |
| | | f | Final condition |
| $M_0$ | Momentum flux $[m^4.s^{-2}]$ | 0 | Initial condition |
| $n_{jet}$ | Number of jets | *Acronyms*: | |
| $n_{vec}$ | Number of vectors | ASTRID: | **A**dvanced **S**odium **T**echnological **R**eactor for **I**ndustrial **D**emonstration |
| $p$ | Static pressure $[Pa]$ | | |
| $Q$ | Mass flow $[m^3.s^{-1}]$ | CEA: | **C**ommissariat aux **E**nergies **A**tomiques et aux **E**nergies **A**lternatives |
| $r$ | Radial jet coordinate | | |
| $S_{jet}$ | Exit surface of a jet $[m]$ | IHX: | **I**nternal **H**eat E**x**changer |
| $s$ | Inter-jets distance $[m]$ | MICAS: | **M**aquette **I**nstrumentée du **C**ollecteur Chaud d'**AS**TRID |
| $T$ | Temperature $[°C]$ | | |
| $U$ | Axial velocity $[m.s^{-1}]$ | SFR: | **S**odium **F**ast **R**eactors |
| $u$ | Velocity $[m.s^{-1}]$ | UCS: | **U**pper **C**ore **S**tructure |
| $V$ | Radial velocity $[m.s^{-1}]$ | | |
| $x$ | Planar jet axial coordinate | *Dimensionless numbers:* | |
| $x_0$ | Horizontal coordinate | $Eu$ | Euler number |
| $X_Z$ | Distance from the jet's exit to its raise $[m]$ | $Fr$ | Froude number |
| | | $Fr_D$ | Densimetric Froude number |
| $z$ | Planar jet coordinate perpendicular to the axial coordinate $x$ | $Pe$ | Peclet number |
| | | $Pr$ | Prandtl number |
| $z_0$ | Vertical coordinate | $Re$ | Reynolds number |

**ACKNOWLEDGMENTS**

The authors acknowledge the support from the CEA and its collaboration with the University of Lorraine, especially the Energy & Theoretical and Applied Mechanics Laboratory. The authors also acknowledge the team of the PLATEAU platform for their support during the experimental campaigns.